\documentclass[11pt]{article}
\usepackage{amsmath,amssymb,color}

\usepackage{amsfonts}

\newcommand{\hoch}[1]{$\, ^{#1}$}

\newcommand{\be}{\begin{equation}}
\newcommand{\ee}{\end{equation}}
\newcommand{\bea}{\setlength\arraycolsep{2pt} \begin{eqnarray}}
\newcommand{\eea}{\end{eqnarray}}

\def\ft#1#2{{\textstyle{\frac{\scriptstyle #1}{\scriptstyle #2} } }}
\def\fft#1#2{{\frac{#1}{#2}}}

\def\0{{\sst{(0)}}}
\def\1{{\sst{(1)}}}
\def\2{{\sst{(2)}}}
\def\3{{\sst{(3)}}}
\def\4{{\sst{(4)}}}
\def\5{{\sst{(5)}}}
\def\6{{\sst{(6)}}}
\def\7{{\sst{(7)}}}
\def\8{{\sst{(8)}}}
\def\sst#1{{\scriptscriptstyle #1}}

\begin{document}

\begin{flushright}
\hfill{CAS-KITPC/ITP-323}
\end{flushright}

\vspace{25pt}
\begin{center}
{\Large {\bf Supersymmetry of the Schr\"odinger and PP Wave
Solutions in Einstein-Weyl Supergravities}}

\vspace{10pt}

Hai-Shan Liu\hoch{1,2} and H. L\"u\hoch{3}

\vspace{10pt}

\hoch{1} {\it Institute for Advanced Physics \& Mathematics,
Zhejiang University of Technology, Hangzhou 310032, China}

\vspace{10pt}

\hoch{2}{\it Zheijiang Institute of Modern Physics\\
Department of Physics, Zhejiang University, Hangzhou 310027, China}

\vspace{10pt}

\hoch{3}{\it Department of Physics, Beijing Normal University,
Beijing 100875, China}

\vspace{40pt}

\underline{ABSTRACT}
\end{center}

We obtain the  Schr\"odinger and general pp-wave solutions with or
without the massive vector in Einstein-Weyl supergravity. The
vector is an auxiliary field in the off-shell
supermultiplet and it acquires a kinetic term in the Weyl-squared super invariant. We study the supersymmetry of these solutions
and find that turning on the massive vector has a consequence of
breaking all the supersymmetry.  The Schr\"odinger and also the
pp-wave solutions with the massive vector turned off on the other hand preserve $\ft14$ of the supersymmetry.

\vfill {\footnotesize Emails: hsliu.zju@gmail.com;\ \ \
mrhonglu@gmail.com}

\thispagestyle{empty}





\newpage

\section{Introduction}

One important application of the AdS/CFT correspondence involves the
construction of gravity duals for non-relativistic field theories.
The non-relativistic conformal symmetry is known as Schr\"odinger
symmetry, which is a symmetry group of the Schr\"odinger equation for free fermions \cite{Hagen:1972pd,Niederer:1972zz} or of fermions at unitarity \cite{Mehen:1999nd}.  Recently, it was proposed that the
corresponding gravity background is a special class of the
cosmological pp-wave solution, namely
\cite{Son:2008ye,Balasubramanian:2008dm}
\begin{equation}
ds^2=\ell^2\Big(-r^{2z} dt^2 + \fft{dr^2}{r^2} + r^2 (-2dt dx +
dy^i dy^i)\Big)\,.
\end{equation}
For $z=1$, it is simply the AdS vacuum with the full Lorentzian
conformal symmetry. For $z=-1/2$, the metric is also Einstein, known as Kaigorodv metric, describing a pp-wave propagating in the AdS background.
It was demonstrated that the isometry of this
metric has the Schr\"odinger symmetry when $z=2$ \cite{Son:2008ye}.
For general $z$, one special conformal transformation is lost
\cite{Balasubramanian:2008dm}. The metric is homogeneous with the
following scaling symmetry
\begin{equation}
r\rightarrow \lambda^{-1} r\,,\qquad t\rightarrow \lambda^z\,
t\,,\qquad x\rightarrow \lambda^{2-z} x\,,\qquad y\rightarrow
\lambda y\,.\label{scalesym}
\end{equation}
The Schr\"odinger solution belongs to the general class of pp-wave solutions. For a generic scaling exponent $z$, the solution can be obtained by introducing a massive vector field in Einstein gravity \cite{Son:2008ye}.  An alternative and well studied geometry that breaks the Lorentz symmetry is the Lifshitz solutions \cite{Kachru:2008yh}. (See also
\cite{Koroteev:2007yp} for an earlier discussion.)

    Massive vector fields are typically absent in usual on-shell
supergravities. However, they can arise in supergravities in the
off-shell formalism as auxiliary fields in the off-shell
supermultiplets.  They are called auxiliary fields because in a two-derivative off-shell supergravity the equations of motion of these fields are algebraic.  However, they can become dynamical when
additional higher-order super invariants are added in the
Lagrangian.  The simplest such a theory is perhaps the Einstein-Weyl
supergravity in four dimensions
\cite{LeDu:1997us,lpsw}. It is the supersymmetric
generalization of the Einstein-Weyl gravity which has been recently
studied for its critical behavior \cite{lpcritical}. Schr\"odinger
and more general pp-wave solutions were demonstrated to exist in
Einstein-Weyl or more general higher-derivative gravities \cite{sol1,sol2,sol3}. Furthermore, it was shown that there exist
new asymptotic AdS and Lifshitz black holes in Einstein-Weyl gravity
\cite{Lu:2012xu}.  Supersymmetric AdS and Lifshitz solutions were
also constructed recently, in which the massive vector plays an
important role for the solutions to be supersymmetric
\cite{Lu:2012am}. These examples show that higher-derivative gravities and supergravities have rich structures for constructing geometric backgrounds that are dual to non-relativistic field theories, as well as the relativistic ones.

    In this paper, we construct the Schr\"odinger and general pp-wave
solutions in Einstein-Weyl supergravity.  Turning on the massive
vector gives rise to new such solutions. We then study the
supersymmetry of these solutions.  We find that the pure
gravitational pp-wave solutions always preserve $\ft14$ of the
supersymmetry.  Turning on the massive vector in these solutions
always breaks the supersymmetry. In contrast, for supersymmetric
Lifshitz solutions in Einstein-Weyl supergravity, the supersymmetry
requires a non-vanishing massive vector \cite{Lu:2012am}.

    The paper is organized as follows.  In section 2, we give a
quick review of Einstein-Weyl off-shell supergravity.  In section 3,
we construct Sch\"odinger solutions with or without the massive vector.
We then study their supersymmetry. In section 4, we construct a general class of pp-wave solutions and
then study the supersymmetry.  In particular we obtain solutions
that describe flows from one Schr\"odinger solution to another with
different scaling symmetries (\ref{scalesym}). We conclude our paper
in section 5.

\section{Einstein-Weyl supergravity}

The field content of the off-shell ${\cal N}=1$, $D=4$ supergravity
consists of the vielbein $e_{\mu}^a$, a massive vector $A$ and a
complex scalar $S + {\rm i} P$, totalling 12 off-shell degrees of
freedom, matching with that of the gravitino $\psi_\mu$.  The
general formalism for constructing a supersymmetric action for any
chiral superfield was obtained in \cite{Binetruy:2000zx}.  For
appropriate choices of superfields, one obtains the actions of the
supersymmetrizations of the cosmological term, the Einstein Hilbert
term and the Weyl-squared terms \cite{LeDu:1997us}.  Adopting the
notation of \cite{lpsw}, the bosonic Lagrangian is given by
\begin{equation}
e^{-1}{\cal L} = R + \ft23 (A^2 - S^2 - P^2) + 4S\,
\sqrt{-\Lambda/3} + \ft12\alpha C^{\mu\nu\rho\sigma}
C_{\mu\nu\rho\sigma} - \ft13\alpha F^2\,,\label{genlag}
\end{equation}
where $C$ is the Weyl tensor and $F=dA$. The supersymmetric transformation rule for the gravitino is given by
\begin{equation}
\delta \psi_\mu = -D_{\mu} \epsilon - \ft{\rm i}6 (2A_\mu -
\Gamma_{\mu\nu} A^\nu)\Gamma_5 \epsilon - \ft16\Gamma_\mu (S + {\rm
i}\Gamma_5 P)\epsilon\,.\label{susytrans}
\end{equation}
The leading-order supergravity with $\alpha=0$ (and also
$\Lambda=0$) was constructed much earlier in \cite{sw,fv}.  The
super invariant associated with the cosmological term $4S\,
\sqrt{-\Lambda/3}$ was introduced in \cite{lpsw}.

The equations of motion for the scalar fields $S$ and $P$ imply that
\begin{equation}
S=3\sqrt{-\Lambda/3}\,,\qquad P=0\,,\label{scalareom}
\end{equation}
and hence they are auxiliary with no dynamical degree of freedom.
The equations of motion for the vector field and the metric are given by
\begin{eqnarray}
0&=&\alpha\,\nabla^{\mu} F_{\mu\nu} + A_\nu\,,\cr 
0&=& R_{\mu\nu}- \ft12 R g_{\mu\nu} + \Lambda\, g_{\mu\nu} -
\ft23\alpha
(F_{\mu\nu}^2 - \ft14 F^2 g_{\mu\nu})\cr %
&&+\ft23 (A_\mu A_\nu - \ft12 A^2 g_{\mu\nu}) -\alpha
(2\nabla^\rho\nabla^\sigma + R^{\rho\sigma})C_{\mu\rho\sigma\nu}\,.
\end{eqnarray}
Thus we see that for $\alpha=0$, the bosonic theory is effectively
Einstein gravity with a cosmological term, with the vector set to
zero by the equations of motion.  On the other hand, if $\alpha$ is
infinity, the theory is conformal gravity with the vector becoming a
Maxwell field in the bosonic sector. It was shown in \cite{lpsw}
that the theory admits a supersymmetric AdS vacuum with the
cosmological constant $\Lambda$.  The linear spectrum on the AdS
background was analyzed.  The gravity modes are identical to those
in Einstein-Weyl gravity studied in \cite{lpcritical}. (See also
\cite{linear1,linear2,linear3,linear4}.) There is a ghost massive
spin-2 mode in additional to the massless graviton.  The mass is
determined by the quantity $\alpha \Lambda$, the product of the
cosmological constant and the coupling of the Weyl-squared term.
There is a critical phenomenon when
\begin{equation}
\alpha \Lambda = \ft32\,,
\end{equation}
for which the massive spin-2 mode disappears and is replaced by the
log mode \cite{lpcritical,sol1,linear2}. Recently, supersymmetric asymptotic AdS and Lifshitz solutions were obtained in \cite{Lu:2012am}.

\section{Supersymmetry of the Schr\"odinger solutions}

\subsection{Schr\"odinger solutions}

Let us consider the following ansatz for the Schr\"ordinger
solutions \cite{Son:2008ye,Balasubramanian:2008dm}
\begin{equation}
ds^2=\ell^2\Big(-r^{2z} dt^2 + \fft{dr^2}{r^2} + r^2 (-2dx dt +
dy^2)\Big)\,,\qquad A=q r^z dt\,,\label{schans}
\end{equation}
where $q$ is a constant.  Note that the scaling symmetry
(\ref{scalesym}) is preserved even though the massive vector field
is turned on.  We find that the equation for the massive vector
implies that
\begin{equation}
q (\ell^2 + \alpha z(z+1))=0\,.
\end{equation}
Thus we see that for non-vanishing $q$, the coupling constant
$\alpha$ is determined fully by the above solution, giving rise to
the Schr\"odinger solution with
\begin{equation}
\Lambda = -\fft{3}{\ell^2}\,,\qquad
\alpha=-\fft{\ell^2}{z(z+1)}\,,\qquad
q=\fft{3(z-1)}{\sqrt2}\,.\label{qschr}
\end{equation}
On the other hand, if the massive vector field is turned off, namely
$q=0$, we have
\begin{equation}
\Lambda =
-\fft{3}{\ell^2}\,,\qquad\alpha=-\fft{\ell^2}{2z(2z-1)}\,.\label{q0sol}
\end{equation}
When $z=1$, the solution is the AdS vacuum; when $z=-1/2$, the
solution is the Kaigrodov metric describing a pp-wave propagating in
the AdS.  Both solutions are Einstein and hence exist for all values
of $\alpha$. 

The scaling exponent $z$ is determined by the product of the
cosmological constant and the coupling of the Weyl-squared super
invariant term.  For each $\alpha \Lambda$, there can be two allowed
$z$, given by
\begin{eqnarray}
q=0:&& z_\pm = \ft14 \pm \ft14\sqrt{1 + \ft{12}{\alpha
\Lambda}}\,,\cr 
q\ne0:&& z_\pm = -\ft12 \pm \ft12\sqrt{1 + \ft{12}{\alpha
\Lambda}}\,,
\end{eqnarray}
It is of interest to note that at the criticality, both has the same $z=1$ root. As we shall
see later, there are more general solutions that describe flows from
one Schr\"odinger solution with $z_-$ to the other with $z_+$.

\subsection{The supersymmetry}

We now investigate the supersymmetry of the Schr\"odinger solutions
we have obtained so far. We do this by studying the Killing spinor
equation $\delta \psi_\mu=0$, namely
\begin{equation}
-D_{\mu} \epsilon - \ft{\rm i}6 (2A_\mu - \Gamma_{\mu\nu}
A^\nu)\Gamma_5 \epsilon - \ft16\Gamma_\mu (S + {\rm i}\Gamma_5
P)\epsilon=0\,.
\end{equation}
To proceed, we choose the following vielbein base
\begin{equation}
e^+ = \ell r^z dt\,,\qquad e^- = -\ft12 \ell r^z dt - \ell r^{2-z}
dx\,,\qquad e^{\bar y} = \ell r dy\,,\qquad e^{\bar r} =
\fft{\ell}{r}dr\,.
\end{equation}
Here the indices $(+,-,\bar y, \bar r)$ are those for the tangent
spacetime. Note that the scaling symmetry (\ref{scalesym}) is
preserved for each vielbein. The resulting spin connection has the
following non-vanishing components
\begin{equation}
\omega^{\bar y}{}_{\bar r}=\fft1{\ell} e^{\bar y}\,,\quad
\omega^+{}_{\bar r}=-\fft1{\ell} e^+ \,, \quad
\omega^+{}_+=-\fft{z-1}{\ell} e^{\bar r} \,,\quad \omega^-{}_{\bar
r} = \fft1{\ell} e^--\fft1{\ell}(z-1)e^+ \,.
\end{equation}
We find that the components of the Killing spinor equation for the
ansatz (\ref{schans}) become
\begin{eqnarray}
\partial_x \epsilon -\ft12 r^{2-z} \Gamma_-(\Gamma_{\bar{r}}
+1)\epsilon &=& 0 \,, \cr 
\partial_r \epsilon + \ft1{2r} (z-1) \Gamma_{+-} \epsilon + \ft
1{2r} \Gamma_{\bar{r}} \epsilon &=& \ft{\rm i}{6r } q
\Gamma_{\bar{r}-} \Gamma_5 \epsilon \,, \cr 
\partial_y \epsilon + \ft12 r \Gamma_{\bar{y}}(\Gamma_{\bar{r}}+1)
\epsilon &=& \ft{\rm i}6 q r \Gamma_{\bar{y}-} \Gamma_5 \epsilon\,,
\cr 
\partial_t \epsilon + \ft 12 r^z \Gamma_{+}(\Gamma_{\bar{r}}+1) \epsilon -
\ft14 (2z-1) r^z \Gamma_{-\bar{r}} \epsilon  - \ft 14 r^z \Gamma_-
\epsilon &=& \ft{\rm i}6 q r^z( \Gamma_{+-} -2)\Gamma_5 \epsilon\,,
\end{eqnarray}
where
\begin{equation}
\Gamma_+^2 =0=\Gamma_-^2\,,\qquad \{\Gamma_+, \Gamma_-\} = 2\,.
\end{equation}
For $q=0$, we find that a Killing spinor exists, given by
\begin{equation}
\epsilon = r^{\fft{z}2} \epsilon_0\,,\qquad
\Gamma_-\epsilon_0=0\,,\qquad \Gamma_{\bar r} \epsilon_0
=-\epsilon_0\,.\label{qeq0ks}
\end{equation}
Thus the solution preserves $\ft14$ of the supersymmetry.  To show
that there are no further Killing spinors, let us examine the
integrability conditions. We find that
\begin{equation}
0=[\partial_t,\partial_y]\epsilon = \ft12 r^{z+1} (z-1) \Gamma_{\bar
y -} \epsilon\,.
\end{equation}
It follows that the Killing spinor must satisfy $\Gamma_{-}\epsilon=0$ for $z\ne 1$.  After imposing this condition, we find that
\begin{equation}
0=[\partial_t, \partial_r]\epsilon=\ft12 r^{z-1}
(z-1)\Gamma_{+}(\Gamma_{\bar r} + 1) \epsilon\,.
\end{equation}
Thus we see that for $z\ne 1$, the Killing spinor must satisfy the
projections given in (\ref{qeq0ks}).

When $q$ is non-vanishing, there is no Killing spinor at all and
hence the solution (\ref{qschr}) is not supersymmetric. This can
be shown as follows.  The integrability condition,
\begin{equation}
0=[\partial_x,\partial_t] \epsilon = -\ft{\rm i}{6} q r^2
(\Gamma_{\bar{r}} + 2)\Gamma_-\Gamma_5 \epsilon\,,
\end{equation}
implies that any would-be Killing spinor must satisfy
$\Gamma_-\epsilon=0$. Imposing this condition, we find that
\begin{equation}
0=[\partial_t,\partial_y]\epsilon = \ft{\rm i}2 q r^{z+1}
\Gamma_{\bar y}\Gamma_5\epsilon\,.
\end{equation}
It is clear that there is no non-vanishing $\epsilon$ that satisfies
the equation.

   It is somewhat surprising that the pure gravitational Schr\"odinger
solutions in Einstein-Weyl supergravity are supersymmetric whilst those with the non-vanishing massive vector  are not.  This is opposite to the Lifshitz solutions where the massive vector is indispensable for supersymmetry \cite{Lu:2012am}.

\section{Supersymmetry of the PP-wave solutions}

\subsection{General PP wave solutions}

The Schr\"odinger solutions belong to the general class of pp-wave solutions which we study in this section.  The ansatz is given by
\begin{eqnarray}
ds^2&=&-\ell^2 \Big( \fft{dr^2}{r^2}+ r^2( -H dt^2 -2dx dt +
dy^2)\Big)\,,\cr 
A&=& \phi\, dt\,,\label{ppans}
\end{eqnarray}
where $H$ and $\phi$ are functions of the coordinates $r, t$ and
$y$.  It becomes the Schr\"odinger solution if we have $H\sim \phi^2 \sim r^{2z}$.  In general the solution is translational invariant only along
the null $x$ direction. Let $\Lambda=-3/\ell^2$.  The full set of
equations of motion reduces to
\begin{eqnarray}
&&\alpha r^4 \phi_{,rr} + 2 \alpha r^3 \phi_{,r} + \ell^2 r^2 \phi +
\alpha \phi_{,yy}=0\,.\label{aeom}\\
&&\fft{r^2}{2\ell^2} \Big(\alpha \Box^2 + (2\alpha + \ell^2)
\Box\Big)H = \ft23 \phi^2 - \fft{2\alpha}{3\ell^2 r^2}
\Big((\partial_y \phi)^2 + r^4 (\partial_r
\phi)^2\Big)\,,\label{eineom}
\end{eqnarray}
where $\Box$ is the Laplacian of the AdS metric ($H=0$) in
(\ref{ppans}):
\begin{equation}
\Box = r^2 \partial_r^2 + 4 r \partial_r +
\fft{1}{r^2}\partial_y^2\,.
\end{equation}
The critical case corresponds to having $\alpha = -\ell^2/2$ and the
resulting equation of $H$ involves only $\Box^2$.

     Let us present the cohomogeneity-one solutions with $H$ and
$\phi$ being functions of $r$ only.  We find
\begin{eqnarray}
\phi &=& q_- r^{-\ft12(1-m)} + q_+ r^{-\ft12(1+m)}\,,\cr H&=& c_0 +
\fft{c_1}{r^3} + \fft{c_-}{m-3} r^{-\ft12 (3-m)} + \fft{c_+}{m+3}
r^{-\ft12 (3+m)}\cr 
&& + \fft{8q_-^2}{9 (3-m)^2}\, r^{-3 + m} +\fft{8q_+^2}{9 (3+m)^2}\,
r^{-3 - m}\,,\label{cohomo1}
\end{eqnarray}
where
\begin{equation}
m=\sqrt{1-\fft{4\ell^2}{\alpha}}\,.
\end{equation}
The solution is valid for general $m$ except for $m^2=9$, corresponding to
the critical point $\alpha = -\ft12\ell^2$.  At the critical point,
the general solution is given by
\begin{eqnarray}
\phi &=& q_- r + \fft{q_+}{r^2}\,,\cr 
H&=&c_0 + \fft{c_1}{r^3} + c_2 \log r + \fft{c_3(1+3\log r)}{r^3} +
\fft{2q_+^2}{81 r^6} + \ft2{27} q_-^2 (3\log r - 4)\log r\,.
\end{eqnarray}
Note that since the equations of motion (\ref{aeom}) and
(\ref{eineom}) do not involve a derivative with respect to $t$,
it follows that the constant coefficients $q_\pm, c_i$ can be
arbitrary functions of $t$.  The metric associated with the term
$c_1$, corresponding to $z=-\ft12$,  is the Kaigorodov metric and it is
Einstein. It was conjectured that its boundary field theory is
certain conformal field theory in the infinite momentum frame
\cite{Cvetic:1998jf}. The coefficient $c_0$ can be removed by a shift of the coordinate $x$. The solution with $q_\pm=0$ was obtained in
\cite{sol1}. It has a linear $\log r$ term, which is characteristic
of $\Box^2\psi=0$ in critical gravity. Interestingly, for
non-vanishing $q_-$, we have a quadratic $\log r$ term, which is absent in pure Einstein-Weyl gravity.

     In general, we can solve the vector equation (\ref{aeom}) by
separation of variables. Letting $\phi=\chi e^{{\rm i} k y}$, we
find
\begin{equation}
\alpha r^4 \chi_{,rr} + 2 \alpha r^3 \chi_{,r} + \ell^2 r^2 \chi -
\alpha k^2\chi=0\,.
\end{equation}
This equation can be solved in terms of Bessel functions
\begin{equation}
\phi=\fft{1}{\sqrt{r}}\Big(\phi_1(t) I(-\ft12m, \ft{k}{r}) + \phi_2
I(-\ft12m, \ft{k}{r})\Big) e^{{\rm i} k y}\,.
\end{equation}
Substituting this into (\ref{eineom}) leads to the general equation
for $H$.  For $\phi=0$, the solution was obtained in \cite{sol1}.
(See also \cite{sol2,sol3}.)

\subsection{From Schr\"odinger to Schr\"odinger}

From the general cohomogeneity-one solution (\ref{cohomo1}), we can
obtain solutions that describe a flow from one Schr\"odinger vacuum
to another when $r$ runs from 0 to infinity.  The scaling symmetry
(\ref{scalesym}) breaks down for general $r$, but recovers at
$r\rightarrow 0$ and $r\rightarrow \infty$, with the different
scaling exponent at the two regions.  For $q_+=0=q_-$, we have
\begin{equation}
ds^2 = -( c_1 r^{2z_+} + c_2 r^{2z_-} ) dt^2 + r^2 (-2dx dt + dy^2)
+ \fft{dr^2}{r^2}\,,
\end{equation}
where
\begin{equation}
z_\pm = \ft14(1\pm m)\,.\label{zpm1}
\end{equation}
In addition, we can have metric that flows from the Kaigorodov
(Schr\"odinger with $z=-1/2$) to a Schr\"ordinger solution:
\begin{equation}
ds^2 = -( \fft{c_1}{r}  + c_2 r^{2z} ) dt^2 + r^2 (-2dx dt + dy^2) +
\fft{dr^2}{r^2}\,,
\end{equation}
where $z$ can be either $z_+$ or $z_-$ in (\ref{zpm1}). If we turn
on $A$, the flow of the Schr\"odinger solutions is given by
\begin{equation}
ds^2 = -( \tilde c_1 r^{2z_+} + \tilde c_2 r^{2z_-} ) dt^2 + r^2
(-2dx dt + dy^2) + \fft{dr^2}{r^2}\,,\quad A= (q_- r^{z_-} +
q_+ r^{z_+}) dt\,,
\end{equation}
where $\tilde c_1,\tilde c_2$ can be read off from (\ref{cohomo1}) and
\begin{equation}
z_\pm = -\ft12(1\pm m)\,.
\end{equation}
Note that we cannot flow from pure gravitational Sch\"odinger
solution to that with non-vanishing $A$.

\subsection{The supersymmetry}

To study the supersymmetry of the pp-wave solutions, we choose the
following vielbein base for the metric (\ref{ppans})
\begin{eqnarray}
e^+=\ell r dt\,,\qquad e^{\bar y}=\ell r dy\,, \qquad e^{\bar
r}=\fft \ell r dr\,, \qquad e^-=-\fft12 \ell r H dt - \ell r dx\,.
\end{eqnarray}
The corresponding spin connection is given by
\begin{eqnarray}
&&\omega^{\bar y}{}_{\bar r}=\fft1{\ell} e^{\bar y}\,,\qquad
\omega^+{}_{\bar r}= \fft1{\ell} e^+ \,, \qquad \omega^-{}_{\bar y}=
- \fft1{2 \ell r} \fft{\partial H}{\partial y} e^+ \,, \cr 
&&\omega^-{}_{\bar r}= - \fft1{2 \ell} r \fft{\partial H}{\partial
r} e^+ + \fft1{\ell} e^- \,.
\end{eqnarray}
Turning off $A$, the Killing spinor equations are given by
\begin{eqnarray}
&&\partial_r \epsilon + \fft{1}{2r} \Gamma_{\bar{r}} \epsilon = 0
\,, \qquad
\partial_y \epsilon + \ft12 r \Gamma_{\bar{y}}(\Gamma_{\bar{r}}+1) \epsilon =0
\,,\qquad
\partial_x \epsilon - \ft12 r \Gamma_-(\Gamma_{\bar{r}} + 1) \epsilon = 0
\,, \cr 
&&\partial_t \epsilon + \fft r2 \Gamma_+(\Gamma_{\bar{r}}+1)
\epsilon - \ft 14 r H \Gamma_-(\Gamma_{\bar{r}}+1) \epsilon - \ft14
\fft{\partial H}{\partial y} \Gamma_{-\bar{y}} \epsilon - \ft14 r^2
\fft{\partial H}{\partial r} \Gamma_{-\bar{r}} \epsilon =
0\,.\label{phi0kseom}
\end{eqnarray}
It is thus clear that a Killing spinor exists provided that
\begin{equation}
\epsilon = r^{\fft{1}2} \epsilon_0\,,\qquad
\Gamma_-\epsilon_0=0\,,\qquad \Gamma_{\bar r} \epsilon_0
=-\epsilon_0\,.\label{phi0ks}
\end{equation}
The question remains whether this is the most general Killing
spinor.  To see this, let us examine the integrability conditions.
We find that
\begin{equation}
[\partial_t,\partial_y]\epsilon = \ft14\Big( (r^3 \partial_r H +
\partial_y^2 H) \Gamma_{\bar y} + r^2 \partial_y\partial_r
H \Gamma_{\bar r}\Big)\Gamma_-\epsilon=0\,.\label{tyint}
\end{equation}
This implies that in general we have $\Gamma_-\epsilon=0$. (The possibility that the terms in bracket vanish will be discussed later.)
This implies from (\ref{phi0kseom}) that
\begin{equation}
\partial_t \epsilon  + \ft12 r \Gamma_+ (\Gamma_r + 1) \epsilon=0
\end{equation}
This equation can be easily solved, giving $\epsilon=\epsilon(r,y) -
\ft12 rt\Gamma_+ (\Gamma_{\bar r} + 1) \epsilon(r,y)$.  Imposing
$\Gamma_-\epsilon=0$ then leads to (\ref{phi0ks}).

If we consider turning on the massive vector, {\it
i.e.}~$\phi(r,y,t)$ is non-vanishing, we find that the Killing
spinor equations are given by,
\begin{eqnarray}
&&\partial_r \epsilon - \ft{\rm i}{6r^2} \phi
\Gamma_{\bar{r}-}\Gamma_5\epsilon + \ft1{2r} \Gamma_{\bar{r}}
\epsilon = 0\,, \cr 
&&\partial_y \epsilon + \ft12 r \Gamma_{\bar{y}}(\Gamma_{\bar{r}} +
1) \epsilon - \ft{\rm i}{6} \phi \Gamma_{\bar{y}-}\Gamma_5 \epsilon
= 0\,, \cr 
&&\partial_x \epsilon -\ft12 r \Gamma_-(\Gamma_{\bar{r}} +1)
\epsilon = 0\,, \cr 
&&\partial_t \epsilon + \ft12 r \Gamma_+(\Gamma_{\bar{r}} + 1)
\epsilon - \ft14 r H \Gamma_-(\Gamma_{\bar{r}} + 1)\epsilon - \ft14
\fft{\partial H}{\partial y} \Gamma_{-\bar{y}} \epsilon - \ft 14 r^2
\fft{\partial H}{\partial r} \Gamma_{-\bar{r}}\epsilon \cr 
&&\qquad\qquad +\ft{\rm i}3 \phi \Gamma_5 \epsilon - \ft{\rm i} 6
\phi \Gamma_{+-}\Gamma_5 \epsilon = 0\,.
\end{eqnarray}
If we impose the condition $\Gamma_-\epsilon=0$, we find that the
following integrability condition
\begin{equation}
0=[\partial_y,\partial_t]\epsilon = -\ft{\rm i}2 r\phi
\Gamma_{y}\Gamma_5 \epsilon - \ft{\rm i}{2} \partial_y \phi
\Gamma_5\epsilon
\end{equation}
It is then straightforward to deduce that no Killing spinor can be
found for $\phi$ that satisfies the equation of motion.

It is worth remarking that the integrability condition (\ref{tyint})
can be satisfied without imposing $\Gamma_-\epsilon=0$. This
requires that
\begin{equation}
H=\fft{1}{r^2} + y^2\,.\label{halfsusysol}
\end{equation}
The resulting metric is given by
\begin{equation}
ds^2=-dt^2 + r^2 ( - y^2 dt^2 - 2dx dt + dy^2) + \fft{dr^2}{r^2}\,.
\label{localads}
\end{equation}
Note that this metric has also the scaling symmetry (\ref{scalesym})
with $z=0$.  In fact it is easy to verify that this metric is
maximally symmetric and hence locally AdS.

After some straightforward calculation, we find that the Killing
spinors that are independent of the null coordinate $x$ are given by
\begin{equation}
\epsilon = \Big(1 - \ft12 y r \Gamma_y (\Gamma_r+1)\Big)
\Big(2\sqrt{r} - \sqrt{\ft{1}{r}}\,\cot t\,\Gamma_-\Big) \Big(\eta_1
+ \sin t\,\eta_2\Big)\,,
\end{equation}
where $\eta_1$ and $\eta_2$ are constant spinors satisfying the
following projection
\begin{equation}
\Gamma_- \eta_1 =0= \Gamma_+ \eta_2\,,\qquad (\Gamma_r +
1)\eta_1=0=(\Gamma_r+1)\eta_2\,.
\end{equation}
Note that neither $\Gamma_-\epsilon$ nor $(\Gamma_{\bar r} +
1)\epsilon$ vanishes, but instead $\Gamma_-(\Gamma_{\bar r} +1)
\epsilon=0$.  It is easy to verify that
\begin{equation}
(\Gamma_{\bar r} + 1)\epsilon =-\fft{1}{2r} \cot t\, \Gamma_-
\epsilon\,.
\end{equation}
This type of Killing spinors was obtained in \cite{kerimo} in
pp-wave solutions with enhanced supersymmetry. The remaining two Killing spinors are dependent
on $x$. It is a linear function of $x$ since it is easy to verify
that $\partial_x^2 \epsilon=0$.  In applying the AdS/CFT
correspondence for Schr\"odinger solution, the null coordinate is
typically treated as periodic and its quantized momentum lifts the mass of a free scalar mode \cite{Son:2008ye,Balasubramanian:2008dm}.  For the
metric (\ref{localads}), treating the coordinate $x$ periodic breaks
half of the supersymmetry.

\section{Conclusion}

In this paper, we have obtained the Schr\"odinger solutions in
Einstein-Weyl Supergravity.  There are two types of such solutions:
the ones which are pure gravitational and those which are supported by the massive vector. The
exponent $z$ of the Schr\"odinger solutions is determined by the
product of the cosmological constant and the coupling of the
Weyl-square super invariant term.  We then demonstrate that the pure
gravitational Schr\"odinger solutions are supersymmetric, preserving
$\ft14$ of the supersymmetry.  The solutions with non-vanishing
massive vector breaks all the supersymmetry.

We then obtain the general pp-wave solutions in Einstein-Weyl
supergravity, which contain the Schr\"odinger metrics as special
solutions.  Furthermore, we have examples that describe flows from one
Schr\"odinger vacuum to another.  We show that the pure
gravitational pp-waves all preserve $\ft14$ of the supersymmetry,
whilst turning on the massive vector field breaks all the
supersymmetry.  This is quite opposite from the Lifshitz solutions
in Einstein-Weyl supergravity.  It was shown that the supersymmetric
Lifshitz solutions in Einstein-Weyl supergravity are all supported
by the massive vector.

Higher derivative gravities and supergravities have rich structures for constructing geometric backgrounds that are dual to both relativistic and non-relativistic field theories. This work shows that non-relativistic field theories from Schr\"odinger backgrounds can be studied in the context of supersymmetry.

\section*{Acknowledgement}

Liu is grateful to KITPC, Beijing, for hospitality during the course of this work. Liu is supported in part by the National Science Foundation of China (10875103, 11135006) and National Basic Research Program of China
(2010CB833000). L\"u is supported in part by the NSFC grant
11175269.


\begin{thebibliography}{99}

\bibitem{Hagen:1972pd}
  C.R.~Hagen,
{\it Scale and conformal transformations in Galilean-covariant field theory,}
  Phys.\ Rev.\ D {\bf 5}, 377 (1972).

\bibitem{Niederer:1972zz}
  U.~Niederer,
{\it The maximal kinematical invariance group of the free Schr\"odinger equation.,}
  Helv.\ Phys.\ Acta {\bf 45}, 802 (1972).

\bibitem{Mehen:1999nd}
  T.~Mehen, I.W.~Stewart and M.B.~Wise,
{\it Conformal invariance for nonrelativistic field theory,}
  Phys.\ Lett.\ B {\bf 474}, 145 (2000)
  [hep-th/9910025].

\bibitem{Son:2008ye}
D.T.~Son, {\it Toward an AdS/cold atoms correspondence: a geometric
realization of the Schr\"odinger symmetry,}  Phys.\ Rev.\  D {\bf
78}, 046003 (2008) [arXiv:0804.3972 [hep-th]].

\bibitem{Balasubramanian:2008dm}
  K.~Balasubramanian and J.~McGreevy,
{\it Gravity duals for non-relativistic CFTs,}  Phys.\ Rev.\ Lett.\
{\bf 101}, 061601 (2008)  [arXiv:0804.4053 [hep-th]].

\bibitem{Kachru:2008yh}
  S.~Kachru, X.~Liu and M.~Mulligan,
{\it Gravity duals of Lifshitz-like fixed points,}  Phys.\ Rev.\ D
{\bf 78}, 106005 (2008)  [arXiv:0808.1725 [hep-th]].

\bibitem{Koroteev:2007yp}
  P.~Koroteev and M.~Libanov,
{\it On existence of self-tuning solutions in static braneworlds
without singularities,}
  JHEP {\bf 0802}, 104 (2008)
  [arXiv:0712.1136 [hep-th]].

\bibitem{LeDu:1997us}
  R.~Le Du,
{\it Higher derivative supergravity in $U(1)$ superspace,}
  Eur.\ Phys.\ J.\ C {\bf 5}, 181 (1998)
  [hep-th/9706058].

\bibitem{lpsw}
  H.~L\"u, C.N.~Pope, E.~Sezgin and L.~Wulff,
{\it Critical and non-critical Einstein-Weyl supergravity,}
  JHEP {\bf 1110}, 131 (2011)
  [arXiv:1107.2480 [hep-th]].

\bibitem{lpcritical}
  H.~L\"u and C.N.~Pope,
{\it Critical gravity in four dimensions,}
  Phys.\ Rev.\ Lett.\  {\bf 106}, 181302 (2011)
  [arXiv:1101.1971 [hep-th]].

\bibitem{sol1}
  M.~Alishahiha and R.~Fareghbal,
{\it $D$-dimensional log gravity,}
  Phys.\ Rev.\ D {\bf 83}, 084052 (2011)
  [arXiv:1101.5891 [hep-th]].

\bibitem{sol2}
  T.~Malek,
{\it Exact solutions of general relativity and quadratic gravity in
arbitrary dimension,}
  arXiv:1204.0291 [gr-qc].

\bibitem{sol3}
  M.~Gurses, T.C.~Sisman and B.~Tekin,
{\it New exact solutions of quadratic curvature gravity,}
  arXiv:1204.2215 [hep-th].

\bibitem{Lu:2012xu}
  H.~L\"u, Y.~Pang, C.N.~Pope and J.~Vazquez-Poritz,
  {\it AdS and Lifshitz black holes in conformal and Einstein-Weyl
  gravities,}
  arXiv:1204.1062 [hep-th].

\bibitem{Lu:2012am}
  H.~L\"u and Z.-L.~Wang,
{\it Supersymmetric asymptotic AdS and Lifshitz solutions in
Einstein-Weyl and conformal Supergravities,}
  arXiv:1205.2092 [hep-th].

\bibitem{Binetruy:2000zx}
  P.~Binetruy, G.~Girardi and R.~Grimm,
{\it Supergravity couplings: a geometric formulation,}
  Phys.\ Rept.\  {\bf 343}, 255 (2001)
  [arXiv:hep-th/0005225].

\bibitem{sw}
  K.S.~Stelle and P.C.~West,
{\it Minimal auxiliary fields for supergravity,} Phys.\ Lett.\  B
{\bf 74}, 330 (1978).

\bibitem{fv} S.~Ferrara and P.~van Nieuwenhuizen,
{\it The auxiliary fields Of supergravity,}
  Phys.\ Lett.\  B {\bf 74}, 333 (1978).

\bibitem{linear1}
S.~Deser, H.~Liu, H.~L\"u, C.N.~Pope, T.C.~Sisman and B.~Tekin, {\it
Critical points of $D$-dimensional extended gravities,}
  Phys.\ Rev.\ D {\bf 83}, 061502 (2011)
  [arXiv:1101.4009 [hep-th]].

\bibitem{linear2}
  E.A.~Bergshoeff, O.~Hohm, J.~Rosseel and P.K.~Townsend,
{\it Modes of log gravity,}
  Phys.\ Rev.\ D {\bf 83}, 104038 (2011)
  [arXiv:1102.4091 [hep-th]].

\bibitem{linear3}
  M.~Porrati and M.M.~Roberts,
{\it Ghosts of critical gravity,}
  Phys.\ Rev.\ D {\bf 84}, 024013 (2011)
  [arXiv:1104.0674 [hep-th]].

\bibitem{linear4}
  H.~Liu, H.~L\"u and M.~Luo,
{\it On black hole stability in critical gravities,}
  Int.\ J.\ Mod.\ Phys.\ D {\bf 21}, 1250020 (2012)
  [arXiv:1104.2623 [hep-th]].

\bibitem{Cvetic:1998jf}
  M.~Cveti\v c, H.~L\"u and C.~N.~Pope,
  Nucl.\ Phys.\ B {\bf 545}, 309 (1999)
  [hep-th/9810123].

\bibitem{kerimo}
  J.~Kerimo and H.~L\"u,
{\it PP-waves in AdS gauged supergravities and supernumerary
supersymmetry,}
  Phys.\ Rev.\ D {\bf 71}, 065003 (2005)
  [hep-th/0408143].

\end{thebibliography}
\end{document}